\documentclass[fleqn,11p,times]{elsarticle}
\usepackage[utf8]{inputenc}

\usepackage{setspace}
\usepackage{graphicx}
\usepackage{graphics}
\usepackage{xcolor}
\definecolor{navy}{rgb}{0.000000,0.000000,0.501961}
\graphicspath{{Images/}}
\usepackage[labelfont=bf]{caption}
\usepackage{subcaption}

\usepackage[ruled,vlined]{algorithm2e}

\usepackage{amsmath}

\usepackage{amsthm}
\makeatletter
\def\els@aparagraph[#1]#2{\elsparagraph[#1]{#2\@addpunct{.}}}
\def\els@bparagraph#1{\elsparagraph*{#1\@addpunct{.}}}
\SetAlgoCaptionSeparator{.}
\RestyleAlgo{algoruled}

\makeatother

\usepackage{float}



\begin{document}

\begin{frontmatter}

\title{Design of A-Star based heuristic algorithm for efficient interdiction in multi-Layer networks}

\vspace{0.4cm}
\author[1,2]{Sukanya Samanta\corref{cor1}}
\ead{susamanta1@gmail.com}
\cortext[cor1]{Corresponding author}

\vspace{0.4cm}

\address[1]{Department of Informatics, Information Science and Electrical Engineering (ISEE), Kyushu University, Fukuoka, 819-0395, Japan}

\address[2]{Department of Systems Innovation, School of Engineering, The University of Tokyo, Tokyo, 113-8654, Japan}

\vspace{0.4cm}

\begin{abstract}
Intercepting a criminal using limited police resources presents a significant challenge in dynamic crime environments, where the criminal's location continuously changes over time. The complexity is further heightened by the vastness of the transportation network. To tackle this problem, we propose a layered graph representation, in which each time step is associated with a duplicate of the transportation network. For any given set of attacker strategies, a near-optimal defender strategy is computed using the A-Star heuristic algorithm applied to the layered graph. The defender's goal is to maximize the probability of successful interdiction. We evaluate the performance of the proposed method by comparing it with a Mixed-Integer Linear Programming (MILP) approach used for the defender. The comparison considers both computational efficiency and solution quality. The results demonstrate that our approach effectively addresses the complexity of the problem and delivers high-quality solutions within a short computation time.

\end{abstract}
\begin{keyword}
\texttt{Resource allocation, Multi-layer time expanded network, Vehicle routing and scheduling, A-Star heuristic algorithm}
\end{keyword}
\end{frontmatter}

%%%%%%%%%%%%%%%%%%%%%%%%%%%%%%%%%%%%%%%%%%%%%%%%%%%%%%%%%%%%%%%%%%%%%%%%%%%%%%%%%%%%%%%

\section{Introduction}
\label{S:1}
In this paper, we aim to develop efficient interdiction strategies for defenders in the context of an escape interdiction problem. We consider a scenario involving multiple defenders and a single attacker, under the assumption that the defenders are only informed about the crime location within a large transportation network. The primary objective of the defenders is to apprehend the attacker before he can escape the city. Given the complexity of the transportation network, our focus lies in devising effective defender strategies against a predefined set of attacker strategies.

The existing literature highlights a wide range of interdiction problems (\cite{samanta2022literature}). In the context of escape interdiction, \cite{conitzer2006computing} demonstrate that exact methods are generally unsuitable due to the NP-hard nature of the problem, which limits scalability because of high computational complexity. To address these limitations, \cite{samanta2021vehicle} propose a simulation-based approach to generate scalable solutions aimed at enhancing security within large transportation networks. Further, \cite{samanta2024assessing} present a resource allocation framework alongside a genetic algorithm (GA)-based metaheuristic to tackle the escape interdiction problem within a simulated environment. Complementing these works, \cite{saito2009discovering} introduce a layered graph model, where each successive layer represents a time step, effectively transforming a complex temporal network problem into a more tractable structure. This layered graph approach has proven valuable for solving transportation network problems in a computationally efficient manner. Motivated by this, we adopt the layered graph concept to address the escape interdiction problem.

Given the challenges posed by limited police resources and the complexity of the transportation network, generating efficient defender strategies remains a formidable task. To the best of our knowledge, this study is the first to apply the A* algorithm within a multi-layer network framework for this problem domain, with an emphasis on developing practical solution methodologies. The proposed A* algorithm determines effective movement strategies for defenders starting from randomly initialized locations. For attackers, we consider a set of randomly generated paths originating from the crime location and ending at randomly selected exit points, each assigned a probability such that the total sums to one.

Through experiments on a sample network, we demonstrate that our approach significantly outperforms the MILP-based defender strategy in terms of both computational efficiency and solution quality, with minimal trade-offs in performance.

The remainder of the paper is structured as follows: Section 2 presents the problem description, Section 3 describes the proposed solution methodology, Section 4 details the benchmarking algorithm, Section 5 discusses the experimental results, and Section 6 concludes the study.

\section{Problem description}
\label{S:2}

We consider an escape interdiction problem involving multiple defenders, denoted by $\overline{D} = \{d_r \mid r \in R\}$, and a single attacker $\overline{A}$. The total number of defenders is $m$, and the set of all defenders is represented by $\overline{D}$. Here, $r \in R = \{1, \dots, m\}$. The defender's pure strategy is denoted by $S$. A pure strategy for the defender consists of $m$ schedules, i.e., $S = \{S^r : r \in R\}$. The attacker has a finite set of possible actions, denoted by $\acute{A}$, which is provided as input. Our objective is to compute the optimal strategy for the defenders using the proposed algorithm. The defenders act jointly, and a complete defender strategy comprises the strategies of all individual defenders. 

The transportation network is represented as a directed graph $G = (V, E)$, where $E$ is the set of directed edges corresponding to roads, and $V$ is the set of nodes representing intersections. There is a set of predefined exit points in the considered network. $v_\infty$ signifies any exit node in the considered network. The escape interdiction scenario starts at time $0$ and concludes at time $t_{\text{max}} > 0$. 

The attacker’s pure strategy is defined by a sequence of states $A = < a_1 = (v^a_0 , 0), . . . , a_j  = (v_j ,  t^a_j ), . . . , a_k = (v_\infty, t^a_k \leq t_{max})>$. Each state $a_j = (v_j , t^a_j)$ indicates that the attacker is located at node $v_j$ at time $t^a_j$. The mixed strategy for the attacker is denoted by $y = <y_A>$, where $y_A$ represents the probability with which the strategy $A$ is played.

Similarly, the state of defender $d_r$ is denoted by the tuple $S^r = (v^r, t^{r,\text{in}}, t^{r,\text{out}})$, indicating that defender $d_r$ occupies node $v^r$ during the time interval $[t^{r,\text{in}}, t^{r,\text{out}}]$.

For a given attacker state $a_j = (v_j, t^a_j)$ and defender state $s^r_i = (v^r_i , t^{r,\text{in}}_i , t^{r,\text{out}}_i)$, an interception occurs if $v^r_i = v_j$ and $t^{r,\text{in}}_i \leq t^a_j \leq t^{r,\text{out}}_i$. If the attacker successfully escapes from the network, the defender receives a utility of $0$; otherwise, the defender receives a utility of $1$.

The utility of the defender is computed using the probability distribution $y$ over $\acute{A}$, as shown in Eq.~(1). Considering the developed defender strategy $S$, the best defender utility is defined as:

\begin{equation}
U_d ( S, y ) = \sum\limits_{ A \in \acute{A} } U_d ( S, A ) \bullet y_A
\end{equation}
%%%%%%%%%%%%%%%%%%%%%%%%%%%

\section{Proposed A-Star algorithm on multi-layer networks}
\label{S:3}
In this escape interdiction problem, we utilize the concept of a multi-layer network (MLN), where a copy of the entire network is created for each discrete time step. The edge lengths in the original network represent time durations, which guide the formation of connections between nodes across different layers. Specifically, based on the edge length, we determine the appropriate layers from which the source and destination nodes of each edge are selected. In this manner, every edge in the original transportation network corresponds to one or more inter-layer edges in the MLN.

We apply the A-Star algorithm on this multi-layer network to derive a near-optimal strategy for the defender. The A-Star function is defined as $f(n) = g(n) + h(n)$, where $g(n)$ represents the exact cost from the defender’s starting node (initial location) to the current node $n$, and $h(n)$ denotes a heuristic estimate of the cost from node $n$ to a goal node. The function $f(n)$ thus combines both the actual and estimated costs to evaluate a node's priority in path planning.

For the attacker, we assume a mixed strategy set which is composed of multiple paths, each starting from a crime location and ending at one of the predefined exit nodes. Each strategy is associated with a probability, and the sum of all such probabilities equals one. To construct the optimal defender strategy, we first assign weights to nodes based on their presence in the attacker's strategies and the associated mixed probabilities.

The exact cost $g(n)$ for each node is computed by summing the mixed probabilities of all attacker strategies in which the node appears. The heuristic value $h(n)$ of a node is estimated by aggregating the mixed probabilities of all future reachable nodes—those that can be reached from the current node and appear in attacker strategies. Hence, for each node in the MLN, the A-Star value is calculated as $f(n) = g(n) + h(n)$.

We define the weight of each node as $(1 - f(n))$. This value is then propagated to all incoming edges of the node. Using these edge weights, we apply Dijkstra’s algorithm to compute the defender’s strategy that yields the maximum probability of successful interdiction (see Algorithm~\ref{alg:a1}). If Dijkstra’s algorithm returns a total probability $P$, we define the defender's utility $U_d$ as $(1 - P)$, which represents the best interdiction probability achievable using the current defender strategy.

\begin{algorithm}[H]
\SetAlgoLined
\textbf{Input:} Crime node as START, all exit nodes as GOAL, Original graph $(G_0)$, Set of attacker strategies with assigned mixed probabilities\;
\textbf{Output:} Defender strategy maximizing probability of interdiction ($U_d$)\;

\textbf{Construction of layered graphs:}\\
\For{$(i = 0; i < t_{\text{max}}; i++)$}{
    Generate a copy of the original graph labeled as $G_i$\;
    Add an edge from GOAL node of $G_i$ to GOAL node of $G_{i+1}$, except for $G_{t_{\text{max}}-1}$\;
}

\textbf{Connect layers based on time-dependent transitions:}\\
\For{each edge in the original graph}{
    $L$ = edge length, $S$ = source node, $T$ = target node\;
    \For{$(j = 0; j < t_{\text{max}} - L; j++)$}{
        Add edge from $S$ in $G_j$ to $T$ in $G_{j+L}$\;
    }
}
\caption{Near-optimal defender strategy design using time-expanded network.}
\label{alg:a1}
\end{algorithm}

\begin{algorithm}[H]
\SetAlgoLined
\textbf{Step 1: Compute exact value $g(n)$ for all nodes:}\\
Initialize $g(n) = 0$ for all nodes in all graphs\;
\For{each attacker strategy}{
    \For{each state $(n, t_a)$ in the current attacker strategy}{
        Exact cost $g(n)$ of node $n$ in graph $G_{t_a}$ = $g(n)$ + $P_{\text{mix}}$, where $P_{mix}$ is the mixed probability of the current attacker strategy \;
    }
}

\textbf{Update exact cost of GOAL nodes:}\\
$min\_t\_index$ = $M$ where $M$ is large integer value \;
\For{each attacker strategy}{
    \For{each state $(n, t_a)$ in the current attacker strategy}{
        \If{$n$ == GOAL node}{
            \If{$min\_t\_index > t_a$}{
                $min\_t\_index$ = $t_a$\;
            }            
        }

\For{$(i = $min\_t\_index$; i < t_{\text{max}} - 1; i++)$}{
    Update the exact value of the GOAL node at $G_{i+1}$ = exact value of the GOAL node at $G_{i+1}$ + exact value of the GOAL node at $G_{i}$ \;
        }
    }
}

\textbf{Update exact cost of all non-interdicted nodes:}\\
\If{Exact cost $g(n)$ of any node == 0}{
    $g(n)$ of that node = $\infty$\;
}

\end{algorithm}

\begin{algorithm}[H]
\SetAlgoLined
\textbf{Step 2: Compute heuristic value $h(n)$ for all nodes:}\\
Set $h(n) = 0$ for all GOAL nodes in $G_{t_{\text{max}} - 1}$\;
\For{$(k = t_{\text{max}} - 2; k \geq min\_t\_index; k--)$}{
    Heuristic cost $h(n)$ of GOAL node at $G_k$ = exact cost ($g(n)$ of GOAL node at $G_{t_{\text{max}} - 1}$) - exact cost of GOAL node ($g(n)$ at $G_k$)\;
    \For{each incoming edge to the GOAL node $n$ in $G_k$}{
        Propagate $h(n)$ as the heuristic cost of the source node of the edge and repeat this same procedure for all incoming edges considering this source node as destination node\;
        }
    }

\textbf{Step 3: Compute $f(n)$ and apply Dijkstra:}\\
\For{each node in the layered graph}{
    Compute $f(n) = g(n) + h(n)$\;
    \If{$f(n) != \infty$}{
        Set weight of node = $(1 - f(n))$\;
    }   
    Assign this weight to all incoming edges of the node\;
}
Assign a weight of $\infty$ to the edges that do not have an assigned weight \;
Apply Dijkstra algorithm to identify defender strategy with minimal path cost $P$\;
Set $U_d = 1 - P$\;

\Return $U_d$\;
\end{algorithm}

\section{MILP benchmarking algorithm}
\label{S:4}

To establish a benchmark, we adopt the MILP-based approach bestDo, developed by \cite{zhang2017optimal}. This method yields the optimal defender strategy by solving a mixed-integer linear program, assuming a predefined set of attacker strategies. The bestDo formulation represents the MILP-based solution for the defender and provides the optimal strategy that maximizes interdiction effectiveness.

The vehicle interdiction problem has been proven to be NP-hard \citep{zhang2017optimal}. When applied to moderately large urban transportation networks with numerous nodes and edges, the MILP formulation incurs significant computational cost in terms of both space and time. The bestDo model aims to maximize the defender's interception capability within a specified time horizon by formulating the optimal movement plan over the network. It constructs a defender path that intercepts the greatest number of attacker paths, thereby maximizing the defender’s utility.

In this formulation, the attacker seeks to escape the city via any exit node $v_\infty$, starting from a designated crime location denoted by $v^0_a \in V$. The state of a defender is represented by the tuple $(v^r, t^{r,\text{in}}, t^{r,\text{out}})$, indicating the defender’s presence at node $v^r$ during the time interval $[t^{r,\text{in}}, t^{r,\text{out}}]$.

In this context, $s^r_{i, v} = 1$ implies that defender $d_r$ is located at node $v$ in the $i^\text{th}$ state of $S^r$. Equation (3) enforces the initial location of $d_r$ at node $v^r_0$ and ensures that the defender occupies exactly one node per state. In Equations (4) and (5), the binary variable $\omega_{r,i,(v,u)}$ indicates whether the defender moves from node $v$ to node $u$ between state $i$ and $i+1$. Equation (6) sets the temporal bounds for the strategy, starting at time 0 and ending at $t_\text{max}$, with $L_\text{max}$ denoting the maximum number of defender states and $k_{r,i}\delta$ representing the time spent at each state. Equation (7) ensures travel between consecutive nodes follows the shortest path.

Equations (8)–(12) define whether an attacker following path $A$ is intercepted, using the binary variable $z_A$. Specifically, $(v^A_j, t^A_j)$ represents the attacker’s $j^\text{th}$ state along path $A$, while $\gamma^{A,j}{r,i}$ indicates whether the attacker is intercepted by defender $d_r$ at that point. The binary variables $\alpha^{A,j}{r,i}$ and $\beta^{A,j}_{r,i}$ represent whether the attacker arrives after the defender and before the defender departs, respectively. A large constant $M$ is used in the big-M formulation to enforce these conditions.

\begin{equation}
max \; - \sum\limits_{ A \in A^{'}} (1 - z_A) y_A
\end{equation}
\begin{equation}
s.t. \qquad s^{r}_{1, v^{r}_0} = 1,  \sum\limits_{ v \in V\setminus \{v_{\infty}\}} s^{r}_{i, v} = 1 \qquad \forall r, i
\end{equation}
\begin{equation}
\omega_{r,i,(v, u)} \leq min(s^{r}_{i, v}, s^{r}_{i+1, u}) \qquad \forall r, i, u, v
\end{equation}
\begin{equation}
\omega_{r,i,(v, u)} \geq s^{r}_{i, v} + s^{r}_{i+1, u} - 1 \qquad \forall r, i, u, v
\end{equation}
\begin{equation}
t^{r, in}_1 = 0, t^{r, out}_{L^{d}_{max}} = t_{max}, t^{r, out}_i = t^{r, in}_i + k_{r, i}\delta\qquad \forall r, i
\end{equation}
\begin{equation}
t^{r, in}_{i+1} = t^{r, out}_{i} + \sum\limits_{ v, u \in V\setminus \{v_{\infty}\}} dist(v, u) \omega_{r,i,(v, u)} \qquad \forall r, i
\end{equation}
\begin{equation}
-M \alpha^{A, j}_{r, i} \leq t^{r, in}_{i} - t^{A}_{j} \leq M (1 - \alpha^{A, j}_{r, i}) \qquad \forall r, i, A, j
\end{equation}
\begin{equation}
-M \beta^{A, j}_{r, i} \leq t^{A}_{j} - t^{r, out}_{i}  \leq M (1 - \beta^{A, j}_{r, i}) \qquad \forall r, i, A, j
\end{equation}
\begin{equation}
\gamma^{A, j}_{r, i} \leq (\alpha^{A, j}_{r, i} + \beta^{A, j}_{r, i} + s^{r}_{i, v^{A}_j})/ 3 \qquad \forall r, i, A, j
\end{equation}
\begin{equation}
\gamma^{A, j}_{r, i} \geq \alpha^{A, j}_{r, i} + \beta^{A, j}_{r, i} + s^{r}_{i, v^{A}_j} - 2 \qquad \forall r, i, A, j
\end{equation}
\begin{equation}
z_A \leq \sum\limits_{j, r, i} \gamma^{A, j}_{r, i} \qquad \forall A
\end{equation}
\begin{equation}
s^{r}_{i, v}, \omega_{r,i,(v, u)}, \alpha^{A, j}_{r, i}, \beta^{A, j}_{r, i}, \gamma^{A, j}_{r, i}, z_A \in \{0, 1\}
\end{equation}
\begin{equation}
k_{r, i} \in Z_{\geq 0}, t^{r, in}_i, t^{r, out}_{i} \in [0, t_{max}]
\end{equation}

To assess the optimality of our proposed approach, we compare the defender utility and computation time of the developed A-Star algorithm with that of the MILP-based bestDo approach.

\section{Results and discussion}
\label{S:5}

In this section, we present the experimental results of the proposed approach. The algorithm is implemented in Python 3.6 and evaluated on a machine equipped with an Intel(R) Core(TM) 3.20 GHz processor and 8 GB of RAM, running a LINUX operating system. The MILP-based bestDo formulation is solved using CPLEX (version 12.8).

\begin{figure}[H]
\centering
\includegraphics[scale=0.25]{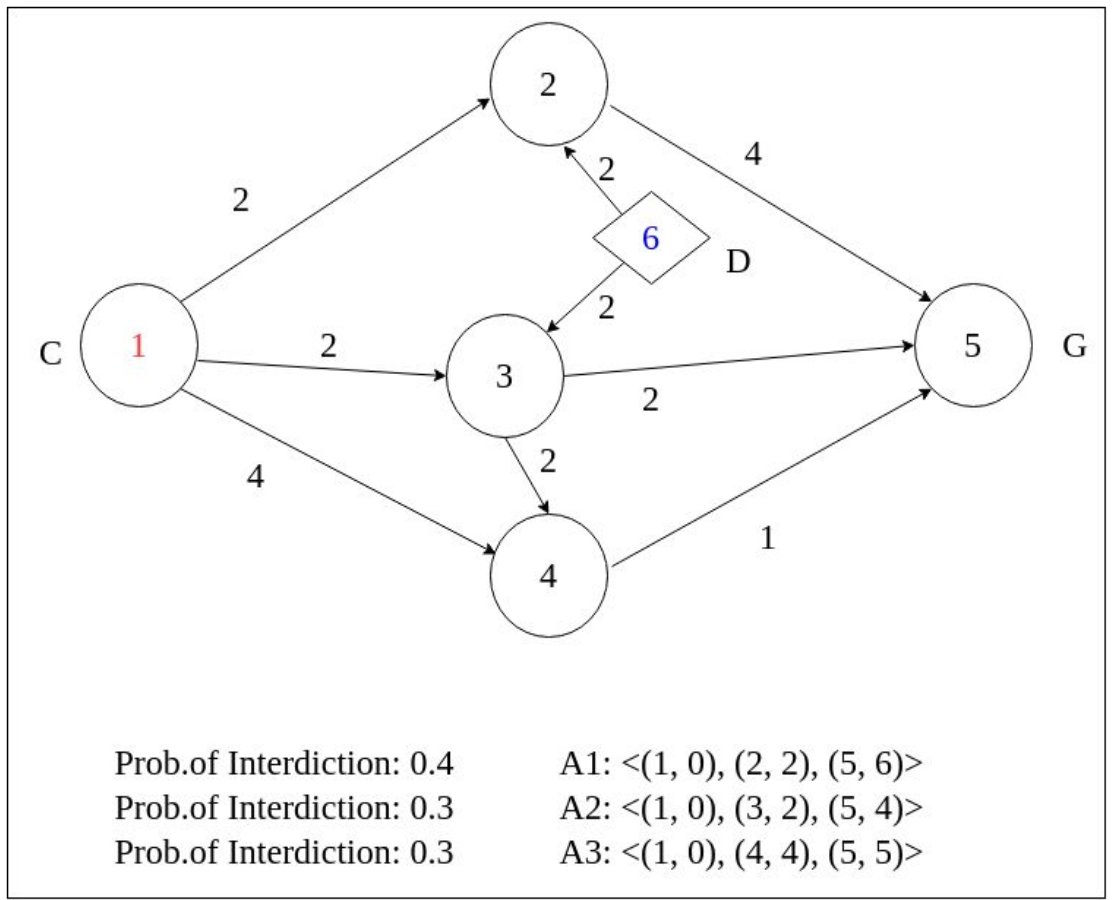}
\caption{Sample network for designing near-optimal defender strategy.}
\label{image-2}
\end{figure}

Fig.~\ref{image-2} illustrates a sample network consisting of 6 nodes, where the police station is located at node 6, the crime scene is at node 1, the time horizon is set as $t_{\max} = 6$, and the designated exit point is node 5. 

\begin{figure}[H]
\centering
\includegraphics[scale=0.36]{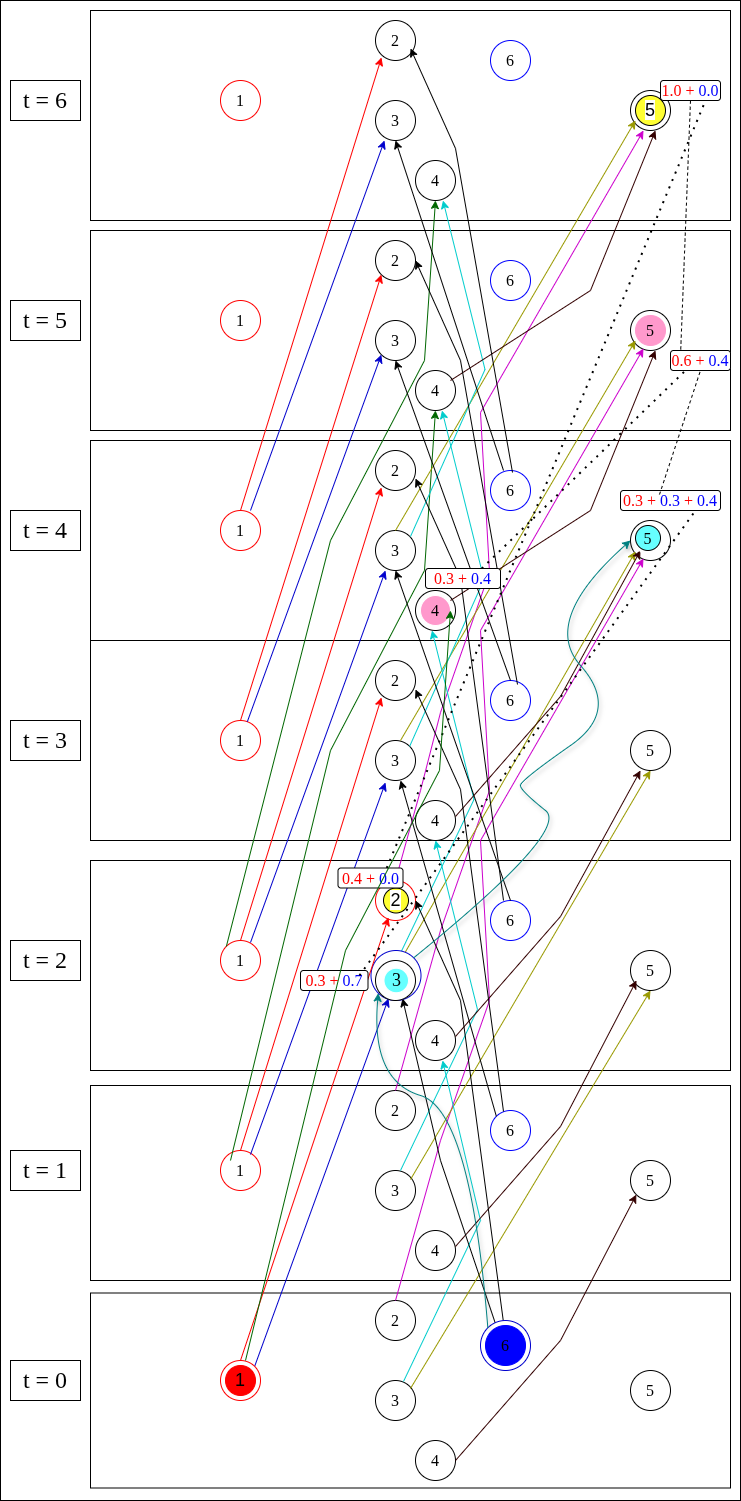}
\caption{Design of a multi-layer network for defender considering a sample network (Fig. \ref{image-2}).}
\label{image-3}
\end{figure}

To generate a near-optimal defender strategy, we apply the A-Star algorithm to the time-expanded network (see Fig.\ref{image-3}). In the time-expanded (multi-layer) network, $0\_6$ denotes node 6 at time $t = 0$. We consider three attacker strategies assigned with a mixed probability. The sum of these mixed probabilities is equal to 1. Nodes belonging to the same attacker strategy are color-coded identically in the multi-layer representation. The crime scene node is colored red, and the initial location of the defender node is colored blue. The resulting defender path is depicted by the green curved lines in the multi-layer network, which follows the path $0\_6 \rightarrow 2\_3  \rightarrow 4\_5 \rightarrow 5\_5 \rightarrow 6\_5$. As shown in Table~\ref{tab:table2}, the proposed approach efficiently constructs a defender strategy that successfully intercepts all attacker paths within the given time limit, considering all ten test cases.

\begin{table}[H]
  \begin{center}
   \caption{\\Defender strategy design using A-Star algorithm.}
    \label{tab:table2}
    \scalebox{0.8}{
    \begin{tabular}{lccccc} % <-- Alignments: 1st column left, 2nd middle and 3rd right, with vertical lines in between
    \hline
      \multicolumn{6}{c}{\textbf{Game Parameters}} \\ \hline
      \multicolumn{6}{c}{\textbf{Network Size: $6 \; Nodes$, Crime Node: $0\_1$, Police Station: $0\_6$, $T_{max}$: $6$}} \\
      \multicolumn{6}{c}{\textbf{Exit Points: 5 [$0\_5, 1\_5, 2\_5, 3\_5, 4\_5, 5\_5, 6\_5$] }}\\
      \multicolumn{6}{c}{\textbf{Input (Attacker Strategies): [[$0\_1, 2\_2, 6\_5], [0\_1, 2\_3, 4\_5], [0\_1, 4\_4, 5\_5]$] }}\\
      \multicolumn{6}{c}{\textbf{Mixed Probabilities: [$0.4, 0.3, 0.3$] }}\\\hline

      \hline
      \textbf{Test Case} & \textbf{Final Defender strategy} & \multicolumn{2}{c}{\textbf{Final Defender Utility}} & \multicolumn{2}{c}{\textbf{Run Time (Sec)}}\\
      \cline{3-4}
      \cline{5-6}
       &  & MILP &  A-Star & MILP &  A-Star \\ \hline
       
       1 & [$0\_6, 2\_3, 4\_5, 5\_5, 6\_5$] & 1 & 1 &  0.337  &  0.041  \\ \hline
       2 & [$0\_6, 2\_3, 4\_5, 5\_5, 6\_5$] & 1 & 1 &  0.337  &  0.042 \\ \hline
       3 & [$0\_6, 2\_3, 4\_5, 5\_5, 6\_5$] & 1 & 1 &  0.337   &  0.039 \\ \hline
       4 & [$0\_6, 2\_3, 4\_5, 5\_5, 6\_5$] & 1 & 1 &  0.337  &  0.040  \\ \hline
       5 & [$0\_6, 2\_3, 4\_5, 5\_5, 6\_5$] & 1 & 1 &   0.337  & 0.039  \\ \hline
       6 & [$0\_6, 2\_3, 4\_5, 5\_5, 6\_5$] & 1 & 1 &  0.337   &  0.048 \\ \hline
       7 & [$0\_6, 2\_3, 4\_5, 5\_5, 6\_5$] & 1 & 1 &   0.337  &  0.040 \\ \hline
       8 & [$0\_6, 2\_3, 4\_5, 5\_5, 6\_5$] & 1 & 1 &  0.337   & 0.048  \\ \hline
       9 & [$0\_6, 2\_3, 4\_5, 5\_5, 6\_5$] & 1 & 1 &  0.337  &  0.044 \\ \hline
       10 & [$0\_6, 2\_3, 4\_5, 5\_5, 6\_5$] & 1 & 1 &  0.337   &  0.041 \\ \hline
    \end{tabular}}
  \end{center}
\end{table}

We compare the defender utility obtained using the proposed A-Star algorithm with that generated by the MILP-based approach, bestDo, introduced by \cite{zhang2017optimal}.

The A-Star evaluation function is defined as $f(n) = g(n) + h(n)$, where $g(n)$ represents the exact cost incurred from the start node (i.e., the initial location of the defender) to the current node $n$, and $h(n)$ denotes the heuristic estimate of the cost from node $n$ to the goal.

We consider three attacker strategies as input:
$a_1 = \langle (1,0), (2,2), (5,6) \rangle$,
$a_2 = \langle (1,0), (3,2), (5,4) \rangle$,
$a_3 = \langle (1,0), (4,4), (5,5) \rangle$,
with corresponding mixed probabilities of $0.4$, $0.3$, and $0.3$, respectively. These mixed probabilities influence the $f(n)$ values assigned to each node.

In Fig.~\ref{image-3}, exact and heuristic costs are indicated in red and blue, respectively. At $t = 2$, node 2 has an exact cost of $0.4$ and a heuristic cost of $0.0$, reflecting the probability mass of future reachable attacker paths. Similarly, at $t = 2$, node 3 has an exact cost of $0.3$ and a heuristic cost of $0.7$. At $t = 4$, node 4 has an exact cost of $0.3$ and a heuristic cost of $0.4$, while node 5 shows an exact cost of $0.3$ and a heuristic cost of $0.7$. The exact cost of $0.3$ for node 5 corresponds to the probability associated with strategy $a_2$, while the heuristic cost of $0.7$ accounts for the future interception potential based on strategies $a_1$ and $a_3$. Continuing, at $t = 5$, node 5 has an exact cost of $0.6$ and a heuristic cost of $0.4$, and at $t = 6$, node 5 has an exact cost of $1.0$ and a heuristic cost of $0.0$.

Both the A-Star and MILP approaches ultimately yield the same defender strategy: $\langle (6,0,0), (3,2,2), (5,4,6) \rangle$, with an interdiction probability of $1.0$. However, the computational time differs: the A-Star algorithm requires approximately 0.04 seconds, while the MILP approach completes in around 0.33 seconds. These test results indicate no performance gap in terms of utility value on this small-scale network, while the A-Star approach demonstrates competitive efficiency in terms of computational time.

\section{Conclusion}
\label{S:6}
In this paper, we introduce an efficient solution methodology for the escape interdiction problem, aimed at addressing security challenges in dynamic crime scenarios. We propose a novel near-optimal routing and resource allocation strategy for defenders operating over a transportation network. Given the attacker's optimal strategy set as input, our approach generates a time-efficient and scalable interdiction strategy by leveraging a multi-layer network representation.

To evaluate performance, we benchmark the results of our proposed A-Star-based method against the adopted MILP approach. A key contribution of this work is the development of a heuristic algorithm that delivers high-quality solutions within significantly reduced computation times, comparable to those of the MILP method.

\textbf{Acknowledgments}
We sincerely thank the members of the Multi-Agent Laboratory at Kyushu University for their valuable discussions and constructive feedback. This research was supported by a project funded through the Grants-in-Aid for Scientific Research program of the Japan Society for the Promotion of Science (JSPS).

{}

\end{document}